\documentstyle[12pt,epsf]{article}
\newcommand{\be}{\begin{equation}}
\newcommand{\ee}{\end{equation}}
\newcommand{\bea}{\begin{eqnarray}}
\newcommand{\eea}{\end{eqnarray}}
\newcommand{\ba}{\begin{array}}
\newcommand{\ea}{\end{array}}

\def\bbox{{\,\lower0.9pt\vbox{\hrule \hbox{\vrule height 0.2 cm
\hskip 0.2 cm \vrule height 0.2 cm}\hrule}\,}}
\newcommand{\dsl}{\pa \kern-0.5em /}

\newcommand{\tg}{\tilde g}

\def\gradhat{ \hat{\nabla } }
\def\bab{ B^{ab} }
\def\psibar {\bar{ \psi} }

\def\psids{{\tilde\psi}}
\def\nablads{{\tilde\nabla}}
\def\tnabla{\tilde\nabla}
\def\tg{\tilde g}

\begin{document}

%%%%%%%%%%%%%%%% title page %%%%%%%%%%%%%%%%%%%%%%%%%%%%%%%%%%%%

\begin{titlepage}
\vfill
\begin{flushright}
hep-th/0206105\\
\end{flushright}

\vfill

\begin{center}
\baselineskip=16pt
{\Large\bf A Positive Energy Theorem for}
\vskip 1.0cm
{\Large\bf Asymptotically deSitter Spacetimes}
\vskip 0.5in
% {\large {\sl }}
% \vskip 10.mm
{\bf David Kastor and Jennie Traschen} \\
%\\[2mm]
\vskip 1cm
%\vfill
{

       Department of Physics\\
       University of Massachusetts\\
       Amherst, MA 01003\\
}
\vspace{6pt}
\end{center}
\vskip 0.5in
\par
\begin{center}
{\bf ABSTRACT}
\end{center}
\vskip 0.3in
\begin{quote}
We construct a set of conserved charges for asymptotically deSitter spacetimes that correspond to asymptotic
{\it conformal isometries}.  The charges are given by boundary integrals at spatial infinity in the flat cosmological 
slicing of deSitter.  Using a spinor construction, we show that the charge associated with conformal time translations is
necessarilly positive and hence may provide a useful definition of energy for these spacetimes.  
A similar spinor construction shows that the charge associated with the time translation Killing vector of deSitter in 
static coordinates has both positive and negative definite contributions.  For Schwarzshild-deSitter the conformal energy we define is given by the mass parameter times the cosmological scale factor.  
The time dependence of the charge is a consequence of a non-zero flux of the corresponding conserved current at spatial infinity.
For small perturbations of deSitter, the charge is given by the total comoving mass density.

\vfill
% \hrule width 5.cm
\vskip 2.mm
\end{quote}
\end{titlepage}

%%%%%%%%%%%%%%%%%%%%%%%%%%%%%%%%%%%%%%%%
\section{Introduction}
In this work we will study the notion of mass in asymptotically deSitter spacetimes.  Recent cosmological
observations indicate that our universe is best described by a spatially flat, Freidman-Robertson-Walker
cosmology with the largest contribution to the energy density coming from some form of `dark
energy'\footnote{See {\it e.g.} \cite{Turner:2002fj} for a recent summary}, matter with an approximate
equation of state $p=w\rho$ with $-1\le w \le -1/2$.  One possibility for the dark energy is a positive
cosmological constant, which is $w=-1$.  In this case, as the universe expands the cosmological constant
will become increasingly dominant, so that in the future our spacetime will tend to increasingly
approximate a region of deSitter spacetime.  These observations have lead to considerable renewed interest
in all things deSitter, including both the classical and quantum mechanical properties of asymptotically
deSitter spacetimes.

Our particular focus in this paper will be to demonstrate positivity properties for the mass of
asymptotically deSitter spacetimes.  At first glance, there are at least three lines of reasoning that
suggest that such positivity properties should not exist for deSitter.  First, Witten's proof of the positive
energy theorem for asymptotically flat spacetimes \cite{Witten:1981mf} and similar results for
asymptotically anti-deSitter spacetimes \cite{Breitenlohner:bm}\cite{Breitenlohner:jf}\cite{Gibbons:aq}
rely on flat Minkowski spacetime and anti-deSitter spacetime respectively being supersymmetric vacuum
states.   DeSitter spacetime is famously not a supersymmetric vacuum state of gauged supergravity. 
Therefore, one would not expect to be able to prove a positive energy theorem.  Second, in global
coordinates the spatial sections of deSitter are closed 3-spheres.  These slices have no spatial infinity at
which to define either asymptotic conditions on the metric, or an ADM-like expression for the mass. 
Rather, as mentioned above for our universe, asymptotic conditions are defined in the past and/or the
future.  Finally, Abbott and Deser have suggested a definition of mass for asymptotically deSitter spacetimes 
\cite{Abbott:1982ff}.  However, they argued that this mass should receive negative contributions from fluctuations outside the deSitter
horizon.

Nonetheless, one further line of reasoning suggests that a positive energy theorem should exist for
deSitter, and we will see that this is in fact the case.  The positive energy theorem for asymptotically flat spacetimes 
may be generalized to
Einstein-Maxwell theory \cite{Gibbons:1982fy} with the result that, if $M$ and $Q$ are the ADM mass and
total charge of the spacetime, then $M\ge |Q|$.  This bound is saturated by the $M=|Q|$ Reissner-Nordstrom
spacetimes and more generally by the well known MP spacetimes \cite{Majumdar}\cite{Papapetrou}, which
describe collections of $M=|Q|$ black holes in mechanical equilibrium with one another.  This bound has a
natural setting in $N=2$ supergravity, and in this context the MP solutions can be shown to preserve
$1/2$ the supersymmetry of the background flat spacetime \cite{Gibbons:1982fy}.  These results closely
parallel those associated with the BPS bound for magnetic monopoles in Yang-Mills-Higgs theory \cite{Witten:mh}.  Hence,
the positive energy bounds in general relativity are regarded as gravitational analogues of BPS bounds. 

It is less well known that multi-black hole solutions also exist in a deSitter background
\cite{Kastor:1993nn}.  The single object solution in this case is a $M=|Q|$ Reissner-Nordstrom-deSitter
(RNdS) black hole.  These spacetimes reduce to the MP spacetimes if one sets the cosmological constant
$\Lambda=0$, and it seems reasonable  to call them MPdS spacetimes\footnote{For $\Lambda>0$ the
physical properties of the MPdS spacetimes depart from those of the MP spacetimes in interesting ways. 
The relation $M=|Q|$ is no longer the extremality condition for RNdS black holes with $\Lambda>0$.  
Rather, it can be shown to be the condition for thermal equilibrium between the black holes and the deSitter
background.  For $M=|Q|$ the Hawking temperatures associated with the black hole and deSitter horizons
are equal.  The solutions with $\Lambda>0$ are also not static.  The black holes follow the
expansion or contraction of the background deSitter universe.}.  It is plausible that the MPdS
spacetimes saturate a BPS bound in the form of a positive energy theorem for asymptotically deSitter
spacetimes.  

It is interesting to ask how the positive energy theorem we present overcomes the three objections stated above?
First, with regard to supersymmetry, we will see that our construction is related to the deSitter supergravity theory
presented in \cite{Pilch:1985aw}.  The super-covariant derivative operator acting on spinors, which we define below, 
coincides with the differential operator in the supersymmetry transformation law for the gravitino field in 
\cite{Pilch:1985aw}.  Although the quantized deSitter supergravity theory is non-unitary, we see that it is nonetheless
useful for deriving classical results.

Second, we must specify an asymptotic region where the mass is to be defined.  As stated above, an asymptotically
deSitter spacetime is specified by conditions at past and/or future infinity.  To define a mass for such spacetimes, we must
consider spatial slices that asymptotically approach one of these regions.  For exact deSitter spacetime, flat cosmological
coordinates, as in equations (\ref{flatslicing}) and (\ref{transform}) below, provide an example of such a slicing.  Figure (\ref{f1})
below shows the conformal diagram for deSitter spacetime, with spatial slices at two different times sketched in.  We see that the point
at the upper left hand corner of the diagram can be reached by going to infinite distance along a spatial slice and can therefore be
regarded as spatial infinity.  We will assume below that the spacetimes under consideration have an asymptotically deSitter region in 
the future and consider a slicing such that the metric approaches that of deSitter spacetime in the flat cosmological slicing.  We will
adopt a pragmatic approach to fall-off conditions, requiring that our constructions be well defined,
for example, for a galaxy in deSitter.

Finally, there is the issue of the non-positivity of the mass for asymptotically deSitter spacetimes defined by Abbott and 
Deser \cite{Abbott:1982ff}.  The construction in reference \cite{Abbott:1982ff} holds generally 
for any class of spacetimes  
asymptotic, with appropriate fall-off conditions, to a fixed background spacetime.  If the background spacetime has isometries, it is 
shown that a conserved charge, defined by a boundary integral in the asymptotic region, can be associated with each of the background
Killing vectors.  For example, the ADM mass of an asymptotically flat spacetime corresponds in this way to the time translation isometry
of Minkowski spacetime.  DeSitter spacetimes are maximally symmetric and therefore also have a maximal number of conserved charges.
A natural choice to call the mass of an asymptotically deSitter spacetime is the charge associated with the time translation Killing vector
for the deSitter background written in static coordinates \cite{Abbott:1982ff}.  However, this Killing vector becomes spacelike outside
the deSitter horizon and correspondingly the mass receives negative contributions from matter or gravitational fluctuations outside the 
deSitter horizon \cite{Abbott:1982ff}\cite{Nakao:1994fj}. Unlike anti-deSitter spacetimes, deSitter spacetimes have no globally timelike Killing vector fields and 
therefore one would not expect to find charges with positivity properties.

DeSitter spacetime does have a globally timelike conformal Killing vector fields (CKV).  If (anti-)deSitter spacetime is viewed as a hyperboloid
embedded in a flat spacetime of the correct signature, the CKVs are simply the projections onto the hyperboloid of the 
translation Killing vectors of the flat spacetime.  For deSitter, time translation symmetry in the embedding spaces yields a globally timelike
CKV.  We will show that there is a conserved charge associated with this CKV,
and that the charge is positive.

This positive, conserved charge, which we denote $Q_\psi$, has a simple interpetation in the perturbative limit.  
Let $\delta\rho$ be the perturbation to the mass density as measured by a comoving cosmological observer.  
We show that $\delta Q_\psi = a(t) \int \delta  \rho$, {\it i.e.} for linear perturbations, the conformal mass at time $t$ is the 
scale factor times the total comoving mass perturbation.  The time dependence of the charge corresponds to a non-zero flux of the 
corresponding conserved current at spatial infinity.  For Schwarzchild-deSitter one finds $Q_\psi =a(t) M$, where
$M$ is the constant mass parameter in the metric.
   
The subsequent sections of the paper are structured as follows.  In section (\ref{basics}), we introduce some necessary elements of (anti-)deSitter geometry.  In section 
(\ref{positiveenergy}) we present a positivity proof, in the manner of Witten \cite{Witten:1981mf}, for the charge $Q_\psi$ defined in terms of
a boundary integral over spinor fields in the asymptotic deSitter region.  
In order to interpret this result, in section (\ref{conservedcharges}) we show that the Abbott and Deser
construction \cite{Abbott:1982ff} can be straightforwardly generalized to associate a conserved charge $Q_\xi$ with {\it any} vector field.  We speculate
that this result is related to Wald's construction \cite{Wald:nt} of a Noether charge associated with an arbitrary spacetime diffeomorphism.  
We show that the charge $Q_\psi$ constructed in section (\ref{positiveenergy}) to identical 
to the charge $Q_\xi$, where the vector $\xi^a$ is a
conformal Killing vector of the background deSitter spacetime.  
We will also discuss the dynamical role of this conformal energy as the Hamiltonian generating 
conformal time evolution in the sense of reference 
\cite{Regge:1974zd}.  In an appendix we give a spinor construction that applies to the 
Killing vectors of deSitter and shows that the exact volume integrand
for the charge has both positive and negative semi-definite contributions\footnote{See, however, reference \cite{Torii:2001} 
where it is argued that the Abbott \& Deser mass nonetheless turns out to be overall positive.}.

In closing the introduction, we note a number of questions suggested by our results, which we will leave for future work.  First is the extension to Einstein-Maxwell theory.  We
would like to check that the MPdS spacetimes of \cite{Kastor:1993nn} do indeed saturate a BPS bound related to the positive energy theorem presented 
here.  A second interesting direction would be to explore the relationship between the conserved charges $Q_\xi$ of section (\ref{conservedcharges}) and Wald's
Noether charges \cite{Wald:nt}.  A third direction would be a more formally complete treatment of asymptotic falloff conditions and a calculation of
the Poisson bracket algebra of conserved charges as in \cite{Regge:1974zd}.  A fourth would be to see if our results are useful in the context of 
the dS/CFT conjecture (see \cite{Strominger:2001pn} and references thereto).

Finally, we would like to comment on the relation of two papers to the present work.  First is an earlier paper of our own \cite{Kastor:1996mj}.  In 
that work, we derived the basic positivity result for asymptotically deSitter spacetimes presented in section (\ref{positiveenergy}) of the present paper.
However, we incorrectly interpreted the spinor boundary term as the charge associated with the time translation Killing vector of deSitter spacetime in static
coordinates, {\it i.e.} the Abbott \& Deser mass.  The more complete presentation here demonstrates the correct interpretation in terms of a new conserved
charge $Q_\xi$ associated with translations in the conformal time coordinate.  Second is a recent paper by Shiromizu et. al. \cite{Torii:2001}.  This paper independently
arrives at the association of the charge $Q_\psi$  with the conformal time translation CKV\footnote{The overall line of reasoning and focus
of reference \cite{Torii:2001} is quite different from the present paper.  In particular positivity of the charge $Q_\psi$ is derived starting from the ordinary positive 
energy theorem in a conformally related asymptotically flat spacetime.}.

\section{Some DeSitter Basics}\label{basics}

In this section, we present a number of properties of deSitter spacetimes that will be useful in following sections.  
In particular, we will focus on the conformal Killing vectors of deSitter and expressions for them in terms of suitably 
defined Killing spinors.  To keep the formulas simple, we will work in $D=4$.  Extension of the results to higher dimensions is
straightforward.  For comparison, we will also present many parallel formulas for anti-deSitter spacetimes.

\vskip 0.2in\noindent{\bf  Carving Out (A)dS from Flat Space}\vskip 0.15in

Start with the embeddings of $4$-dimensional
$dS$ and $AdS$ as hyperboloids respectively in $(4+1)$ and $(3+2)$
dimensional Minkowski spacetime.  These can be studied together by
writing the embedding equation for (A)dS with cosmological constant $\Lambda = 3\kappa /R^2$ as
\be\label{hyperboloid}
-(X^0)^2 + (X^1)^2+(X^2)^2+(X^3)^2+\kappa (X^4)^2 =R^2,
\ee
where $\kappa=+1$ for deSitter and $\kappa =-1$ for anti-deSitter.  It will be useful below to write the
five dimensional flat metric as
\be\label{flat}
      ds_{5}^2= \kappa dR^2 +R^2 k_{a b}dx^a dx^ b .
\ee
where $x^a$ with $a, b=0,1,2,3$ are arbitrary coordinates on $(A)dS$ and
$g_{a b}=R^2 k_{a b}(x^c)$ is the $(A)dS$ metric with radius $R$.
We will also frequently choose flat cosmological coordinates $(t,\vec x)$ for deSitter, 
in which the $dS$ metric takes the form
\be\label{flatslicing}
ds^2=-dt^2+ a(t)^2 d\vec x\cdot d\vec x, \qquad a(t)=e^{Ht},
\ee
where the Hubble constant $H=R^{-1}$.
These cosmological coordinates are related to the flat embedding
coordinates in equation (\ref{hyperboloid}) via the relations
\be\label{transform}
t=R \log \left( {X^0+X^4\over R}\right), \qquad x^k= {RX^k\over X^0+X^4},
\qquad k=1,2,3 .
\ee

\begin{figure}[!ht]
\begin{center}
{\epsfysize=1.75in \epsfbox{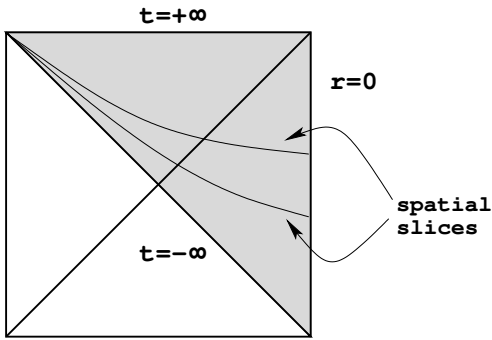}}
\end{center}
\caption{The conformal diagram for deSitter spacetime with the region covered by the flat cosmological coordinates of equations
(\ref{flatslicing}) and (\ref{transform}) shaded and spatial
slices at two different times are indicated. }
\label{f1}
\end{figure}

\vskip 0.2in\noindent{\bf Conformal Killing Vectors}\vskip 0.15in

The isometry groups
of $dS$ and $AdS$ are the Lorentz groups $SO(4,1)$ and $SO(3,2)$ of the respective flat embedding spacetimes.  
However, the
groups of conformal isometries for both $dS$ and $AdS$ (as well as for
$(3+1)$ dimensional Minskowski spacetime) are the same\footnote{In a
general the spaces $(A)dS_p$ share the conformal group $SO(p,2)$ with
$(p-1,1)$ dimensional Minkowski spacetime.} $SO(4,2)$.  One can think of
the extra conformal isometries as arising from the translational Killing
vectors of the flat $5$ dimensional embedding spacetime.   The CKV's of
$(A)dS$ are simply the projections of these vectors onto the $(A)dS$
hyperboloids. This can be demonstrated easily starting from the form of the 5 dimensional flat metric in 
equation (\ref{flat}).  Let
$\xi^{(A)}=\partial/\partial X^A$.   The vectors $\xi^{(A)}$ satisfy
$\nabla^{(5)}_B\xi^{(A)}_C=0$.
After transforming to the $(R,y^a)$ coordinates adapted to the
hyperboloid, we then have in particular
\bea\label{gradfive}
0  &  =  &  \nabla^{(5)}_a\xi^{(A)}_b  \\
& = & \nabla_a \xi^{(A)}_b -\Gamma^R_{ab}\xi^{A}_R,
\eea
where $\nabla_\alpha$ is the $(A)dS$ covariant derivative operator.
The relevant Christoffel symbol $\Gamma^R_{\alpha\beta}$ is simply related
to the extrinsic curvature of the embedding
and is found from the metric (\ref{flat}) to be
$\Gamma^R_{\alpha\beta}=-{\kappa\over R}g_{\alpha\beta} $.  For the $(A)dS$ space of radius $R$, we then have
\be\label{ckv}
\nabla_a \xi^{(A)}_b= -{\kappa \over R}g_{ab}\xi^{(A)}_R
\ee
Projection of the five dimensional vectors $\xi^{(A)}$ onto the hyperboloid
amounts to dropping the five dimensional radial component
$\xi^{(A)R}$.  Therefore, we continue to use the symbol $\xi^{(A)}$ to
denote the four dimensional projected vector.  Comparison of equation
(\ref{ckv}) with the conformal Killing equation
$\nabla_a\xi_b+\nabla_b\xi_a=2 f g_{ab}$ shows
that the four dimensional vectors
$\xi^{(A)}$ are CKVs with corresponding conformal factors
\be\label{factors}
f^{(A)}=-{\kappa \over R}\xi^{(A)}_R
\ee
One can also derive a useful expression for the gradients of the conformal
factors $f^{(A)}$.  Along with equation (\ref{gradfive}), one has
\bea\label{gradfivetwo}
0  &  =  &  \nabla^{(5)}_a\xi^{(A)}_R \\
& = & \nabla_a \xi^{(A)}_R -\Gamma_{a R}^b\xi^{(A)}_b.
\eea
Plugging in $\Gamma_{a R}^b = {1\over R}\delta_{ab}$ and using equation (\ref{factors}) then gives
\be\label{specialprop}
\nabla_a f^{(A)}=- \kappa H^2\xi^{(A)}_a
\ee
We see that the gradients of the conformal factors are simply proportional
to the CKV's themselves.  

Finally, we will need the explicit forms of the $dS$ CKV's
$\xi^{(A)}$ in the flat cosmological coordinates defined in
(\ref{transform}),  %
\bea
\xi^{(0)} & =  &\left(\cosh (Ht) +{r^2 H^2 \over 2}e^{Ht}
\right)\left({\partial\over\partial t}\right) -
H x^ke^{-Ht}\left({\partial\over\partial x^k}\right) \nonumber \\
\xi^{(4)} & =  &\left(-\sinh (Ht) +{H^2 r^2\over 2} e^{Ht}
\right)\left({\partial\over\partial t}\right) -
H x^ke^{-Ht}\left({\partial\over\partial x^k}\right) \label{CKV}\\
\xi^{(k)} & =  & -x^ke^{Ht} \left({\partial\over\partial t}\right) +
{1\over H}e^{-Ht}\left({\partial\over\partial x^k}\right)\nonumber
\eea
We note that the particular linear combination of CKV's, $\zeta =\xi^{(0)}-\xi^{(4)}$, is given simply by  
\be\label{conftime}
\zeta =  e^{Ht} 
\left({\partial\over\partial t}\right) 
\ee
and is everywhere timelike and future directed.  The CKV
$\zeta$ generates translations in the conformal time
parameter $\eta$ defined by $d\eta=\exp{(-Ht)}dt$.  
We will see below that it is a conserved charge associated with $\zeta$ that satisfies a positivity relation.	

\vskip 0.2in\noindent{\bf Killing Spinors}\vskip 0.15in

We now turn to spinor fields in $(A)dS$.  The Killing spinors of the flat 
embedding spacetimes lead to Killing spinors in $(A)dS$ that are constant with respect to the 
projection of the 5 dimensional covariant derivative operator onto the $(A)dS$ hyperboloid.  
Denoting this projected derivative operator by $\breve\nabla_a$, we have
\be
\breve\nabla_a\eta
= \nabla_a\eta +{1\over 2}\omega_a^{\,\,\hat b\hat
R}\gamma_{\hat b\hat R}\eta,
\ee
where hats denote orthonormal frame indices.
The necessary component of the spin connection for the metric (\ref{flat}) is given by
$\omega_a^{\,\,\hat b\hat R}={\kappa\over R} e_{\ a}^{\hat b}$, where
$e_{\ a}^{\hat b}$ are the components of the vierbein for the $(A)dS$
metric $g_{ab}$.  This yields the expression
\be\label{super}
\breve\nabla_a\eta=\nabla_a\eta+{\kappa \over 2R}\gamma_a\gamma_{\hat
5}\eta,
\ee
Both $dS$ and $AdS$ inherit four complex solutions to
$\breve\nabla_a\phi=0$, which we call Killing spinors, from the flat space embeddings.  
The explicit solutions for the Killing spinors of $dS$ in the flat cosmological 
coordinates of (\ref{flatslicing}) are given by
\be\label{spinorsol}
\eta (\vec x,t)= \left( 1- {H\over 2} x^k\gamma_k (\gamma^{\hat
5}+\gamma^{\hat t} )
\right )\left(e^{Ht\over 2} \alpha _-  + e^{-Ht\over 2} \alpha _+ \right)
\ee
where  $\alpha_\pm$ are constant spinors satisfying the conditions
\be\label{project}
(\gamma^{\hat 5}\mp \gamma^{\hat t})\alpha _\pm = 0. 
\ee
Below, we will find it convenient to work with another form of the derivative operator.  If one takes
$ \eta = {1\over\sqrt{2} } (1+ c \gamma^{\hat 5} ) \psi $
with $c^2=-\kappa$, then one finds that 
\be
\breve\nabla_a\eta={1\over\sqrt{2} } (1+c\gamma^{\hat 5})\hat\nabla_a\psi
\ee
where the new derivative operator $\hat\nabla_a$ is given by
\be
\hat\nabla_a\psi=\nabla_a\psi +{c\over 2R}\gamma_a\psi .
\ee
With $c=1$ for $AdS$, $\hat\nabla_a$ is the super-covariant derivative operator that comes from $N=2$ 
gauged supergravity as given in {\it e.g.} \cite{Romans:1992nq}.  With $c=i$ for $dS$, $\hat\nabla_a$ is 
the supercovariant derivative operator arising in the classical deSitter supergravity theory discussed in 
\cite{Pilch:1985aw}.

Bilinears in the Killing spinors give linear combinations of the (C)KV's for $(A)dS$ in the following way.  
If $\psi$ satisfies $\hat\nabla_a\psi =0$, one can show that the vector $\xi^a=-\bar\psi\gamma^a\psi$ satisfies
\be\label{gradxi}
\nabla_a\xi_ b=
\left\{  \begin{array}{ll}
{i\over R}g_{a b}\bar\psi\psi , & {\mbox dS}\\
& \\
-{1\over R} \bar\psi\gamma_{a b}\gamma_{\hat 5}\psi, & {\mbox AdS}
\end{array}\right .
\ee
We see that for $AdS$, the right hand side is anti-symmetric and $\xi^a$ is therefore a Killing vector.  
For $dS$, the right hand side is proportional to the metric, and $\xi^a$ is therefore a conformal Killing vector 
instead.    It is also straightforward to show that the vectors $\xi^a=-\bar\psi\gamma^a\psi$ are everywhere future 
directed and timelike.  Specializing to $dS$ in the cosmological coordinates (\ref{flatslicing}), one finds that the 
particular CKV $\zeta^a$ generating translations in the conformal time coordinate $\eta$ is obtained by taking $\alpha_+=0$ in 
equation (\ref{spinorsol}).

The situation is reversed if one looks at the vectors
$\chi^a=-\bar\psi \gamma_{\hat 5}\gamma^a \psi$, which satisfy
\be\label{gradxitwo}
\nabla_a\chi_ b=
\left\{  \begin{array}{ll}
-{i\over R} \bar\psi\gamma_{ab} \gamma_{\hat 5}\psi, & {\mbox dS}\\
& \\
{1\over R} g_{a b} \bar\psi\psi , & {\mbox AdS}
\end{array}\right .
\ee
The vectors $\chi^a$ are therefore Killing vectors for $dS$ and conformal
Killing vectors for $AdS$.  Note that unlike the case of flat spacetime, 
the Killing vector one obtains from the Killing spinors are not irrotational, {\it i.e.} 
the antisymmetric piece $\nabla_{ [a } \xi_{ b ]}$ is nonzero.

\section{A Positive Energy Theorem for DeSitter}\label{positiveenergy}

\subsection{Derivation of the Spinor Identity}

In this section, we derive a Gauss's law identity for spinor fields in $4$-dimensional, asymptotically (A)dS spacetimes using the 
derivative operator $\hat\nabla_a$.  Following the covariant approach of reference \cite{nester}, we start by defining the Nester form 
$B^{ab}= \bar\psi \gamma^{abc}\hat\nabla_c \psi$.
It is then straightforward to show that the divergence of twice the real part of $B^{ab}$ can be written as
\bea\label{spinoridentity}
\nabla_a\left( B^{a b}+ B^{a b *}\right)& 
= & -\left (G^ b_{\  c}+\Lambda g^ b_{\  c}\right)
\bar\psi\gamma^ c\psi
      +2 \overline{(\hat\nabla_a\psi)}
\gamma^{a b c}(\hat\nabla_ c\psi)  \\
& & +{1\over R} (c-c^*)\left ( \bar\psi\gamma^{a b}(\hat\nabla_a\psi)
      - \overline{(\hat\nabla_a\psi)}\gamma^{a b}\psi\right), \nonumber
\eea
where $G_{ab}$ is the Einstein tensor and we have made use of the identity  
$\gamma ^{abc} \nabla _a \nabla _c
\psi = -{1\over 2}
 G^{bc} \gamma_c\psi.$
Recalling that $c=i$ for dS and $c=1$ for AdS, we see that the coefficient of last term 
in equation (\ref{spinoridentity}) vanishes for AdS.

Now choose a spacelike slice $\Sigma$ with timelike unit normal $n^a$.  
The spacetime metric can then be written as $g_{ab}=-n_an_b +q_{ab}$, with $q_{ab}$ the metric on $\Sigma$.
Stokes theorem implies that a $2$-form field $A_{ab}$ satisfies
\be\label{stokes}
 \int _V (\nabla_b A^{bc}) n_c =
 -{1\over 2}\int _{\partial V} ds_{ab}A^{ab},
\ee
where the spatial volume $V$ lies within $\Sigma$ and has boundary $\partial V$.
For a spinor field $\psi$ define the quantity $Q_\psi$ by the surface integral
\bea\label{definecharge}
Q_\psi  & = & -{1\over 16\pi} \int _{\partial V}ds_{ab} \left( B^{a b}+ B^{a b *}\right)\\
& = & -{1\over 16\pi} \int _{\partial V} ds_{ab} \left( \bar\psi
\gamma^{abc}(\hat\nabla_c \psi)
- (\hat\nabla _c \bar\psi )\gamma^{abc} \psi\right )\label{surfaceform}
\eea
Stokes theorem together with the identity (\ref{spinoridentity}) then imply that $Q_\psi$ may be 
alternatively written as a volume integral
\bea\label{spinoridtwo}
 Q_\psi & = 
& {1\over 8\pi}\int _V \left [  \left( G_{a b}+\Lambda g_{a b}\right) \xi^a n^ b
+  q^{ij}(\hat\nabla_i\psi^\dagger)  \hat\nabla_j\psi  -
      (\gamma^i\hat\nabla_i\psi )^\dagger (\gamma^j\hat\nabla_j\psi) \right .\\
 & &\left . -{1\over R} (c-c^*)  \left(\psi^\dagger (\gamma^j\hat\nabla_j\psi)
- (\gamma^j\hat\nabla_j\psi)^\dagger \psi \right) \right ],
\eea
where $\xi^a=-\bar\psi\gamma^a\psi$ and the indices $i,j$ are raised and lowered with the
spatial metric $q_{ij}$.
If we now impose that the spinor field satisfy the Dirac-Witten equation
\be\label{dirac}
\gamma^j\hat\nabla_j\psi =0,
\ee
then on solutions to the Einstein equations $ G_{ab}+\Lambda g_{ab} =8\pi T_{ab}$ one has 
the relation
\be\label{positive}
 Q_\psi
= {1\over 8\pi} \int _\Sigma \left ( 8\pi T_{ab} \xi^a n^b +  q^{ij}(\hat\nabla_i\psi)^\dagger
\hat\nabla_j\psi \right )
\ee

The second term is manifestly non-negative, vanishing only if $\hat\nabla_i\psi=0$.
Since the vector field $\xi^a=-\bar\psi\gamma^a\psi$ is by construction everywhere future directed and timelike, 
the first term will also be non-negative provided that the stress-energy tensor satisfies the dominant energy condition (see {\it e.g.} 
\cite{Wald:rg}).   Subject to this condition on the stress-energy tensor, we have then proven that  
\be\label{bound}
Q_\psi\ge 0.
\ee
For anti-deSitter spacetimes this is the standard result of \cite{Gibbons:aq}.  
The charge $Q_\psi$ in this case is a linear combination of the conserved charges defined in reference \cite{Abbott:1982ff},
corresponding to the Killing vectors of anti-deSitter spacetimes.
For deSitter spacetimes, equation (\ref{bound}) is a new result.  
The nature of the charge $Q_\psi$ in deSitter remains to be explored.
Since $\xi^a=-\bar\psi\gamma^a\psi$ is a conformal Killing vector in this case, we expect that the charge $Q_\psi$ is a conserved
quantity related to the conformal Killing vectors of deSitter spacetimes.
Indeed,  since $B^{ab}$ is antisymmetric, the divergence of left hand side
of (\ref{spinoridentity}) is identically equal to zero,
\be\label{spinorcurrent}
\nabla _b J^b \equiv \nabla _b \nabla_a \left( B^{ab}+ B^{ab*}\right) =0
\ee
Hence $Q_\psi$ is the volume integral of the time component
of a conserved current.  We further discuss the interpetation of $Q_\psi$ in section (5).

The above analysis can be repeated using the more general two form
$A^{ab}=\bar\psi (\gamma^{\hat 5})^p
\gamma^{abc}\hat\nabla_c\psi$.
For $p=0$ this is the same as the case we have been analyzing, and
$Q_\psi$ depends on
the CKV's of deSitter. For $p=1$, $Q_\psi$ depends on the KV's of deSitter.
This construction is summarized in the
Appendix.  However, in the case of the KV's we show that
the $\gradhat\psi$ contribution to the volume integrand is not
positive definite.
Therefore even if there are no matter sources, $T_{ab}=0$, the
charge generated by
a Killing vector does not appear to be positive definite\footnote{See however the argument in reference \cite{Torii:2001} that the Abbott \& Deser
mass is nonetheless positive semi-definite.}. 
Alternatively, one could consider the KV case for perturbations off deSitter. However,
none of the KV's is timelike everywhere, so again the integrand is not
positive definite.

\subsection{\bf Evaluation of $Q_\psi$ in the Asymptotic Limit}\label{asymptotic}

We now want to rewrite the boundary integral expression for the charge $Q_\psi$ (\ref{surfaceform}) in terms of the asymptotic behavior of the
metric.  We begin by writing the spacetime metric in the asymptotic region as $g_{ab} =\tilde g_{ab} + h_{ab}$.  Here $\tilde g_{ab}$ is the 
deSitter metric\footnote{In the following sections, tilde's will denote deSitter quantities.  For example, the Dirac matrices 
$\tilde\gamma^a$ satisfy the algebra $\{\tilde\gamma^a,\tilde\gamma^b\}=2\tilde g^{ab}$.}.  We will work with flat cosmological
coordinates, in which the asymptotic deSitter metric takes the form (\ref{flatslicing}).
The rate of falloff we require for the metric perturbation $h_{ab}$ in the asymptotic region is discussed below.  
We assume that in the asymptotic region, the spinor field has the form $\psi=\psids + \delta\psi$, where $\psids$ is a deSitter Killing spinor
and $\delta\psi$ is falling off to zero.

We are seeking to write the boundary integral expression for $Q_\psi$ in terms of $h_{ab}$.  The most conceptually straightforward 
way to do this would be to start with a general metric perturbation $h_{ab}$ satisfying the asymptotic conditions, solve the Dirac-Witten
equation subject to the boundary condition that $\psi$ approach a Killing spinor, and then substitute into the expression (\ref{surfaceform})
for $Q_\psi$.  In the asymptotically flat case \cite{Witten:1981mf}, this calculation leads to the result 
$Q_\psi = -P_a\xi^a$, where $\xi^a$ is the future directed timelike Killing vector built out of the asymptotic Killing spinor.  Positivity of $Q_\psi$ then
implies that the ADM $4$-momentum $P^a$ is future directed, timelike or null.

In the present case, this approach 
proves to be quite cumbersome due to the fact that the
extrinsic curvature of the background is nonzero.  Instead we will follow the
 method used in reference \cite{Lee:aq}.  We will also 
restrict our analysis to spinor fields $\psi$  that approach Killing 
spinors which are
eigenfunctions of $\gamma ^{\hat t}$. These have
$\alpha_+=0$ in equation (\ref{spinorsol}).  As noted above, 
the vector $\xi^a=-\bar\psi\gamma^a\psi$ formed from
these spinors is the conformal time translation CKV.  The main result 
of this section is the expression (\ref{finally}) for the charge 
$Q_\psi$
corresponding to these spinors.  The derivation of (\ref{finally}) is 
somewhat lengthy and the reader may want to simply
skip ahead to this equation to focus on the results.

Following reference \cite{Lee:aq}, we begin by splitting the 
derivative operator
$\hat\nabla_a$ into a background piece and a perturbation 
$\hat\nabla_a =\hat\nablads_a + (\gradhat_a - \hat\nablads_a)$
and writing the Nester form $B^{ab}$ as
\be\label{dividebt}
       \bab [\gradhat ] = \bab [\hat\nablads] + \bab [\gradhat - 
\hat\nablads ] .
\ee
We first show that the background part of the boundary integral vanishes,
\be\label{vanish}
\int _{\partial \Sigma} ds_{ab} \bab [\hat\nablads]=0.
\ee
As indicated above, we take the Killing spinor $\psids$ to be given by
$\psids  = {1\over \sqrt{2}}(1- i\gamma ^{\hat 5}) e^{Ht/2}\alpha_+$. 
With this choice
the boundary integral (\ref{vanish}) is found to depend only on the 
$1/r$ terms in $\delta\psi$.
This choice of $\psids$ is an 
eigenspinor\footnote{The Killing spinor
constructed from $\alpha_+$ is not an
eigenspinor of $\gamma ^{\hat t}$ and hence the argument below would not
imply finiteness of $Q_\psi$ in this case.} of $\gamma ^{\hat t}$, satisfying
$\gamma ^{\hat t} \psids =-i \psids$.
The argument now proceeds by the
following steps.

\vskip 0.15in\noindent{\bf 1:} Show that the boundary integral 
vanishes if $\delta \psi$ also satisfies
$\gamma ^{\hat t} \delta \psi =-i \delta \psi$.
%eigenfunction of $\gamma ^{\hat t}$ with eigenvalue $-i$.
\hfill\break
{\bf 2:} Determine consistent falloff conditions on the metric 
perturbation $h_{ab}$, such that the
Einstein constraint equations are satisfied.\hfill\break
{\bf 3:} Show that to order $1/r$, $\delta\psi$ satisfies the 
requirement of step 1.

\vskip 0.15in\noindent {\bf Step 1:}
We want to show that equation (\ref{vanish}) holds, if
$\gamma ^{\hat t} \delta \psi =-i \delta \psi$.
To leading order in $1/r$ we have,
\be
B^{ab} (\hat\nablads )=\bar\psids \gamma^{abc} (\nablads_c +i{H\over 
2} \gamma _c)
\delta \psi
\ee
Integration by parts together with the relation $\hat\nablads_a\psids 
=0$ then yields
\be\label{parts}
\int _{\partial \Sigma} ds_{ab}  \bar\psids \gamma ^{abc}
\nablads _c \delta \psi =
i{H\over 2} \int _{\partial \Sigma}  ds_{ab} \bar\psids \gamma _c \gamma
^{abc} \delta \psi.
\ee
Using this result, the surface integral of $B^{ab} [\hat\nablads ]$ 
in (\ref{vanish}) can  then be written as
$2iH \int _{\partial \Sigma} ds _{tj} \bar\psids \gamma^t \gamma ^j 
\delta \psi$.
The integrand then vanishes if both $\psids$ and $\delta\psi$ are 
eigenspinors of $\gamma^{\hat t}$ with the
same eigenvalue.
The rest of the argument below shows that $\delta\psi$ is indeed such 
an eigenspinor to order $1/r$.

\vskip 0.15in\noindent {\bf Step 2:} We must now discuss the falloff 
conditions on $h_{ab}$ in the asymptotic region. We require that
the rate of fall off is general enough to include solutions to the Einstein constraint
equations corresponding to localized stress energy sources with non-zero monopole moments.
That is, we want the resulting positive mass theorem to include the case of a galaxy in
deSitter\footnote{See also
Shiromizu et al \cite{Torii:2001} for an alternative discussion of 
falloff conditions in asymptotically deSitter spacetimes, which 
arrives
at the same conditions.}.  

Schwarzschild-deSitter spacetime is useful as a reference, since we can solve the
Dirac-Witten equation in this case. In static 
coordinates the metric is
\be\label{SdS}
ds^2= - F(R)dT^2 +{1\over F(R)}dR^2 +R^2 d\Omega^2,\qquad 
F(R)=1-{2M\over R} -{\Lambda\over 3} R^2.
\ee
The Killing vector $\partial/\partial T$ is, of course, only timelike 
in the region between the deSitter and black hole horizons.
Transforming to cosmological coordinates, the metric (\ref{SdS}) 
becomes the McVittie metric \cite{mcvittie}
\be\label{mcVittie}
ds^2 = -{(1- {M\over 2ar} )^2\over  (1+ {M\over 2ar} )^2}\ dt^2 +
a^2 (1+ {M\over 2ar} )^4 \ \delta _{ij} dx^i dx^j ,\qquad a(t)=e^{Ht}
\ee
Then the equation $\gamma^j\hat\nabla_j\psi =0$ 
is solved exactly
for the McVittie metric by the spinors
\be
\psi=\Omega^{-2}\psids,\qquad \Omega=1+ {M\over 2ar}
\ee
where $\psids$ is a Killing spinor of the background deSitter spacetime. Subsitituting
$\psi$ into the boundary term, one explicitly calculates $Q_{\psi}=a(t) M$. 
Both the metric and the spinor fields fall off like $r^{-1}$, as one expects. However,
note that simple power counting suggests that the boundary integral (\ref{parts}) diverges:
$\tilde\psi$ is constant, $\delta\psi$ goes like $ r^{-1}$, and the area element
goes like $r^2$. The fact that (\ref{parts}) is zero (rather than infinity) is due to the
orthogonality properties of the spinors involved, which follow from the fact that
$\psi$ is an eigenfunction of $\gamma^{\hat t}$.

Turning to the general case, let $\Sigma$
be a spatial slice  with unit normal $n^a$, spatial metric 
$q_{ij}$, extrinsic curvature $K^{ij}$, and let
$\rho =T_{ab}n^a n^b$ be the matter density. 
To determine appropriate falloff conditions we use the Hamiltonian constraint equation
\be\label{hamiltonian}
R^{(3)} + K^2 -K ^{ij}K_{ij}  -2\Lambda =16\pi \rho
\ee
If $\rho$ is a compact source, then in the far field the spatial metric $q_{ij}$ satisfies a Poisson-type
equation, and hence the perturbation to $q_{ij}$ vanishes like $1/r$. Therefore we
require that in the asymptotic region,the spatial metric has the form 
\be\label{spacefalloff}
q_{ij}=\tilde q_{ij}+ {\cal O}(1/r).
\ee
Let $k^{ij} =K^{ij}-\tilde K ^{ij}$ be the perturbation to the 
extrinsic curvature. In the case of the McVittie metric (\ref{mcVittie}), the extrinsic curvature 
is given by $K^{ij} = Hq^{ij}$,
which exactly balances the cosmological constant term in the
Hamiltonian constraint equation (\ref{hamiltonian}).
In general, we find it judicious to
let $K^{ij} = Hq^{ij} + l^{ij}$, so that the perturbation to the 
extrinsic curvature becomes $k^{ij} =
H (q^{ij}-\tilde q^{ij}) + l^{ij}.$
Equation (\ref{spacefalloff}) implies that the $3$-dimensional scalar curvature at infinity vanishes as
$R^{(3)} \sim{\cal O}(1/r^3)$. Substituting into the
 constraint equation then implies that  
%$\tilde k^{ab}\sim r^{-n+1}$.
%
\be
l^{ij}\sim {\cal O}(1/r^3)
\ee
The order
%$O(r^{-n+3})$
$O(1/r)$ corrections to the extrinsic curvature
enter only from the $Hq^{ij}$term. The perturbation $l^{ij}$ must fall off
more rapidly because the background extrinsic curvature is nonzero\footnote{In $n$ spacetime dimensions the 
fall off conditions become $\delta q_{ij}\sim r^{-n+3}, l^{ij}\sim r^{-n+1}$.}. Note 
that the trace of the extrinsic curvature is given by
%$K= (n-1)H +\tilde k$.
$K= 3H +l$. 
The Einstein momentum constraint does not add any new information.

\vskip 0.15in\noindent {\bf Step 3:} We now show that $\gamma ^{\hat 
t} \delta \psi =-i \delta \psi$ to order $1/r$.
The spinor field $\psi $ must solve the Dirac-Witten equation 
$\gamma^k\hat\nabla_k\psi=0$.
In detail this is
\be
\gamma ^k \partial _k \psi
+{1\over 4} \omega _k ^{\ \hat a\hat b}\gamma^k\gamma _{\hat a\hat 
b}\psi +{3\over 2}i H  \psi
=0,
\ee
where
$\omega _c ^{\ \hat a \hat b} =e ^{\hat a d} (\partial _c e ^{\hat b}_d
-\Gamma ^e _{cd} e^{\hat b} _e)$
is the spin connection.  We choose coordinates that $h_{tj}=0$. This means
that the normal to $\Sigma$ is  asymptotically in the direction of ${\partial\over \partial t}$
  One then finds $\omega _k ^{\ \hat t \hat i}
=K_{\ k} ^i$ and hence
\be
\gamma ^k\omega _k ^{\ \hat t \hat i} \gamma _{\hat t \hat i} = 3H
+l.
\ee
The $\omega_i ^{jk}$  components of the spin connection vanish in the background. 
The Dirac-Witten equation for $\psi$ becomes
\be\label{pertdirac}
    \gamma ^k \partial _k \psi +{1\over 4} \omega _k ^{\ ij}\gamma^k 
\gamma _{ij}\psi +
    {3\over 2}i H (1-i\gamma ^{\hat t} )\psi +{1\over 2}l\psi =0
\ee
We want to show that (\ref{pertdirac}) has solutions 
which are eigenspinors of $\gamma ^{\hat t}$, at least through the
first correction $\delta \psi$.
If this is true, then the third term in equation (\ref{pertdirac}) vanishes.
Assuming that this is the case, now
multiply equation (\ref{pertdirac}) by $\gamma^{\hat t}$ and rewrite it as a
differential equation for $\gamma ^{\hat t}\psi $.  The first two 
terms change sign, but
the last one does not.  However,
$l \sim 1/r^3$ while $\omega_k^{\ \hat i\hat j} \sim 1/r^2$.
The $l$ term is then of higher order.

Therefore to leading order, we want a solution to
\be\label{pertdiractwo}
     \gamma ^k \partial _k \delta\psi =
   -{1\over 4} \omega _k ^{\ ij}\gamma ^k\gamma _{ij}\psids ,
\ee
such that $\delta\psi$ satisfies
$\gamma ^{\hat t}\delta \psi =-i \delta \psi$.  Let $N_A$ be the 
eigenspinors of $\gamma ^{\hat t}$ with
eigenvalue $-i$, and $P_A$ be the eigenspinors with eigenvalue $+i$, 
with $A=1,2$ in four dimensions.
Choose the indexing such that $\psids = \exp(Ht) N_1$.
Multiplying by a spatial Dirac matrix flips the eigenvalue of 
$\gamma^{\hat t}$, so that
that $(\gamma ^j N_A ) =C^j_{AB} P_B$, where
$C^j_{AB}$ is a constant matrix.
Now expand the perturbation $\delta\psi$ to the spinor as $\delta\psi 
=F_B N_B$,
where the $F_B$ are unknown functions to be solved for.
In the differential equation (\ref{pertdiractwo}) for $\delta\psi$ , we have
\be \omega _k^{\ ij}\gamma ^k\gamma _{ij} =
   -\Gamma ^n _{kl}\gamma ^k \gamma ^l _{\ n} = \alpha _i \gamma ^i,
\ee
where the $\alpha _i$ are known functions in terms of derivatives of $h_{ij}$.
Substituting this into equation (\ref{pertdiractwo}) we than have
\be
[C^k_{BC} \partial _k
F_B + e^{Ht} \alpha _i C^i_{1C}] N_C =0.
\ee
This is a set of two
first order PDEs for the
two unknown functions $F_B$, and therefore one expects that 
generically there is a solution.

This completes the argument that the integral $\int _{\partial 
\Sigma} ds_{ab} B^{ab}
[\gradhat ^{dS} ]$ vanishes.  Note that at next order in powers of 
$1/r$ the $l$ term
does contribute to the differential
equation (\ref{pertdirac}), and therefore it is not possible to 
find solutions
which are eigenspnors of $\gamma ^{\hat t}$ everywhere in the volume 
$\Sigma$. However, things
work out just right to have the $B^{ab} (\hat{\tilde\nabla})$ 
contribution vanish at infinity.
Equation (\ref{definecharge}) for the charge $Q_\psi$ now reduces to
\be
  Q_\psi =  -{1\over 16\pi  }\int ds_{ab}\left ( B^{ab} [\gradhat - 
\hat\nablads ] +  B^{ab*} [\gradhat -
\hat\nablads]\right )
\ee
The difference in derivative operators acting on the spinor field 
$\psi$ can then be written as
\be\label{diff}
       (\nabla _a -\nablads_a )\psi
={1\over 4}g_{bc}\, {\cal C}^c _{da}\gamma ^{bd}\psi \quad ,
\ee
with
${\cal C}^a _{bc}= g^{ad} (\nablads _b h_{cd }+ \nablads _c h_{bd } 
-\nablads _d h_{bc})$ and $h_{ab} =g_{ab} -\tilde g_{ab}$.
Substitution then yields an explicit expression for the charge $Q_\psi$
\bea\label{finally}
Q_\psi  = & {1\over 16\pi} \int ds_{ab}
[ \xi ^a (\nablads_d h^{db}-\nablads^{b} h) - \xi ^b (\nablads_d 
h^{da}-\nablads^{a} h) \\
  & \ +\xi ^c (\nablads^{b} h_c ^{\ a} -\nablads^{a} h_c^{\ b} ) ] \nonumber
\eea
where as before $\xi ^a =- \bar{\tilde\psi} \gamma ^a \tilde\psi$ and 
$h=\tilde g^{ab}h_{ab}$.
This is a familiar expression from the asymptotically flat case.  If 
$\tilde\nabla_a$ were instead the flat spacetime
covariant derivative operator and $\xi^a$ a time translation Killing 
vector, then equation (\ref{finally}) would be the expression for
the ADM mass.

\section{CKV's and Conserved Charges}\label{conservedcharges}

\subsection{General Construction}

In section (\ref{positiveenergy}) we constructed charges $Q_\psi$ associated with the conformal Killing
vectors of the background deSitter spacetime.  These charges are conserved.
In particular for the conformal time translation CKV we showed
that $Q_\psi\ge 0$.  This result raises the question, what is the physical significance of these charges?
One usually associates conserved charges with symmetries.  However, this
construction involves the conformal 
symmetries.  In this section, we will see that for asymptotically (anti-)deSitter spacetimes, there exist conserved 
charges corresponding to the background CKV's as well as those corresponding to KV's.

Our result follows from an extension of the methods of reference \cite{Abbott:1982ff}.
The Abbott and Deser construction begins with 
spacetimes that are asymptotic at infinity to a fixed background spacetime, which has some number of
Killing vectors.  For each background Killing vector, they
showed that there exists a conserved current and further that the 
corresponding conserved charge, obtained by integrating the time component of the current over a spatial slice, can be 
re-expressed as a boundary integral at spatial infinity.  The construction in \cite{Abbott:1982ff} is covariant in nature.
We note that these results may also be obtained using the Hamiltonian formalism for general relativity, as a special case of the integral
constraint vector (ICV) construction of reference \cite{Traschen:1985bp} (see also \cite{Abbott:1988ff}).
Here we will follow the covariant approach.  % We will show that the conserved charge In this section 

Consider a spacetime with metric $g_{a b}$ and stress energy tensor $T_{a b}$ that satisfies the Einstein
equations with cosmological constant $\Lambda$,
\be\label{einstein}
R_{a b}-{1\over 2}g_{a b}R + \Lambda g_{a b}= 8\pi T_{a b}.
\ee
Assume that the spacetime is asymptotic at spatial infinity to a fixed background spacetime
with metric $\tilde g_{a b}$ and vanishing stress energy\footnote{The Hamiltonian approach of 
\cite{Traschen:1985bp} allows for background stress-energy tensor $\tilde  T_{a b}\neq 0$. This involves 
introducing vector fields more general than KV's and CKV's, which are the ICV's mentioned above.}  
that also solves the Einstein equations with
cosmological constant $\Lambda$.  Define the tensor $h_{a b}$ to be the difference between the
spacetime metric and the fixed background metric.  
\be
h_{a b}=g_{a b}-\tilde g_{a b}.
\ee
Note that $h_{a b}$ is not
assumed to be small in the interior of the spacetime, but must
vanish at an appropriate rate near spatial infinity (see Section 3.2).  

The Abbott and Deser construction \cite{Abbott:1982ff} proceeds by expanding the curvature of $g_{a b}$ in powers of $h_{a b}$. 
The Einstein equations (\ref{einstein})  are rewritten keeping terms linear in $h_{a b}$ on the
left hand side and moving all the nonlinear terms $G^{(NL)} _{a b}$ to
the right hand side of
the equation, giving  
\be\label{linearized}
R_{a b}^{(L)}-{1\over 2}\tilde g_{a b}R^{(L)}-\Lambda
h_{a b}=  8\pi T_{a b}- G^{(NL)} _{a b}
\ee
where $R_{a b}^{(L)}$ and $R^{(L)}$ are the terms linear in $h_{a b}$ in
the expansions for the Ricci curvature and scalar curvature of $g_{a b}$.
Collecting all the nonlinear terms and the matter
stress-energy into the quantity ${\cal T}_{a b}=8\pi T_{a b}- G^{(NL)} _{a b}$,
equation (\ref{linearized}) can now be processed into the useful form
\be\label{useful}
{\cal T}^{ac}=\tilde\nabla_b\tilde\nabla_d K^{abcd} + {1\over 2} \tilde R^c_{\ bde}K^{abde},
\ee
where indices have been raised using the background metric and $\tilde\nabla_a$ is the covariant derivative operator for the
background metric $\tg_{ab}$.  The tensor $K^{abcd}$ is defined by
\be\label{superpotential}
K^{abcd}={1\over 2} \left [ \tg^{ad} H^{bc} +\tg^{bc}H^{ad}
-\tg^{ac}H^{bd}-\tg^{bd}H^{ac}\right ],
\ee
where $H^{a b}=h^{a b}-{1\over 2}\tg^{a b}h$, and has the same symmetries as the Riemann
tensor. 

The Bianchi identity for the Einstein tensor implies that $\tilde\nabla_a{\cal T}^{a b}=0$.
It then follows that if $\xi^a$ is a Killing vector of the background metric, then the current 
\be
{\cal J}^a={\cal T}^{a b}\xi_ b
\ee
is conserved with respect to the background derivative operator, 
\be\label{current}
\tilde\nabla_a {\cal J}^a=0.
\ee
A conserved charge $Q_\xi$ is now obtained by integrating the normal component of the current
over a spatial slice $\Sigma$ with respect to the background volume element,
\be
Q_\xi=\int_\Sigma  d^3x\sqrt{-\tilde g} \ {\cal J}^a n_a.
\ee
Abbott and Deser \cite{Abbott:1982ff} then further show that the charge $Q_\xi$ can be 
re-expressed as an integral over the boundary $\partial\Sigma$ of the spatial slice $\Sigma$ at
spatial infinity.

We now show how to extend these results to include the CKVs of background (anti-)deSitter spacetimes.  In fact we will
derive a considerably more general result.  
Take any vector field $\xi^a$, not necessarily a background KV or CKV, and contract $\xi^a$ with 
both sides of equation (\ref{useful}). After some algebra it follows that
\bea\label{good}
{\cal T}^{ab}\xi_b & =  & \tilde\nabla_c\left ( (\tilde\nabla_d K^{acbd})\xi_b
-K^{ad bc}\tilde\nabla_d\xi_b\right)  \\
&& +\left( K^{ac bd}\tilde\nabla_d\tilde\nabla_c +{1\over 2}
K^{ae cd}\tilde R^b_{\ ecd}\right)\xi_b \nonumber
\eea
By judiciously moving terms from the right hand side to the left, equation (\ref{good}) can be put in a Gauss's law form
\be\label{gauss}
{\cal C}^a=\tilde\nabla_b {\cal B}^{ab},
\ee
where ${\cal B}^{ab}$ and ${\cal C}^a$ are given by
\bea
{\cal B}^{a b} & = & (\tilde\nabla_c K^{a bd c})\xi_d
-K^{abdc}\tilde\nabla_{[c}\xi_{d]}\label{surfacecharge}\\
{\cal C}^a & = & {\cal T}^{ab}\xi_b + \tilde\nabla_c\left( K^{adbc}
\tilde \nabla _{(d}\xi_{b)}\right)
+\left ( K^{acbd}\tilde\nabla_d\tilde\nabla_c +{1\over 2}
K^{aecd}\tilde R^b_{\ ecd}\right )\xi_b \label{volumecharge},
\eea
The symmetries of $K^{abcd}$ imply that ${\cal B}^{ab}$ is an antiysymmetric tensor. Therefore (\ref{gauss})
 implies that the vector field ${\cal C}^a$ is divergenceless with respect to the background derivative operator,
$\tilde\nabla_a {\cal C}^a=0$.  If $\xi^a$ is a background Killing vector, then the second and third terms in ${\cal C}^a$ vanish, 
and ${\cal C}^a$ reduces to
the current ${\cal J}^a$ defined above.  Equation (\ref{gauss}) is then the result of Abbott and Deser \cite{Abbott:1982ff}.
More generally, however, the Gauss's law identity (\ref{gauss}) holds, and hence also the conservation law 
$\tilde\nabla_a {\cal C}^a=0$, 
holds for any vector field $\xi^a$.  This result may seem surprising.  However, we should recall that Wald \cite{Wald:nt} has shown that
any vector field, acting as a generator of diffeomorphisms, gives rise to a conserved current and corresponding Noether charge.
We expect that our result has an interpretation in this context.

Taking a spatial surface $\Sigma$ with unit timelike normal $n^a$ and boundary $\partial\Sigma$, 
define the quantity $Q_\xi$ by the surface integral
\be\label{adm}
Q_\xi= {1\over 8\pi}\int_{\partial\Sigma}ds_{ab}{\cal B}^{ab}.
\ee
Stokes theorem and equation (\ref{gauss}) then imply that $Q_\xi$ can also be written as the volume integral
\be
Q_\xi= {1\over 8\pi}\int_\Sigma \sqrt{\tilde g}{\cal C}^a n_a .
\ee
The form of the boundary integral (\ref{adm}) does not depend on the presence, or absence, of a cosmological 
constant and it therefore holds, in particular for asymptotically flat spacetimes.  In this case, plugging in the
translational Killing vectors of Minkowski spacetime for $\xi^a$ yields the usual expressions for the components of
the ADM 4-momentum.  Turning to asymptotically deSitter spacetimes, the expression for the Abbott \& Deser mass \cite{Abbott:1982ff}
is obtained by inserting the time translation Killing vector of deSitter spacetime 
in static coordinates.  

In Schwarzschild-deSitter, the charge for the CKV $\zeta = a(t){\partial \over \partial t}$ 
is $Q_\zeta= a(t)M$. 
 One may wonder about the fact that $Q_\zeta$ depends on time $t$, since 
we have shown above that it is a conserved charged.  The time dependence comes about because 
there is a nonzero flux of the spatial components of the current ${\cal C}^k$ through spatial infinity. 
One can explicitly compute the flux of ${\cal C}^k$ for Schwarzchild-deSitter,
and verify that it equals $\partial _t Q_\zeta =H a M$.

\subsection{Equivalence of $Q_\psi$ and $Q_\xi$ for CKV's}

Now consider asymptotically deSitter spacetimes, and let $\xi^a$ be one of the conformal Killing vectors of the background deSitter
spacetime.  The main result of this section is to demonstrate that the charge $Q_\xi$, in this case, is related to the charge 
$Q_\psi$ constructed for deSitter CKV's in section (\ref{positiveenergy}).   We will check this both for the surface intergal and the 
volume integral expressions for the two charges agree.  For the surface integrals, this is straightforward.  Plugging a deSitter 
CKV into the boundary integral expression for $Q_\xi$ (\ref{adm}), the resulting expression can easily be put in the form of the 
boundary integral for $Q_\psi$ given in equation (\ref{finally})
with the identification $\xi^a=-\bar{\tilde\psi}\tilde\gamma^a\tilde\psi$.  In particular, the term proportional to the shear $\tilde\nabla_{[b}\xi_{d]}$
of the CKV vanishes by virtue of equation (\ref{gradxi}).

Having checked that the boundary integral expressions for $Q_\xi$ and $Q_\psi$ agree, it then follows that the volume integral expressions must also agree.
However, it is interesting to check this explicitly, in order to gain some intuition into the conserved charges associated with CKV's.
Of course, it is not possible (or at least not plausible) for a general spacetime to
 demonstrate directly that the volume  integrands for $Q_\xi$ and $Q_\psi$ agree 
to all orders in $h_{ab}$. We will work only to linear order
in $h_{ab}$.

Even at linear order $h_{ab}$, checking agreement of the volume integrals for $Q_\xi$ and $Q_\psi$ is a more complicated task then checking 
agreement of the surface integral expressions.  To see why, note
that the charges $Q_\xi$ and $Q_\psi$, regarded as surface integrals, depend
 respectively only on the behavior of the vector field $\xi^a$ and the
spinor field $\psi$ in the asymptotic region.  There are then infinitely many ways extend $\xi$ and $\psi$ in the interior that keep the corresponding 
charges fixed.  These are essentially gauge degrees of freedom.  In the present case, we have specified that $\xi^a$ will be taken to be a deSitter CKV throughout the interior.  However, we have not 
yet specified how to extend the spinor field $\psi$ into the interior.  There are two natural choices.
  One choice is to simply take $\psi$ to be one
of the deSitter Killing spinors $\tilde\psi$ throughout the spacetime.  Then
 the boundary expression for $Q_\psi$ is still given by (\ref{finally}) .  
This has the disadvantage that the spinor will not satisfy the Dirac-Witten equation (\ref{dirac}) and hence the positivity of $Q_\psi$ is not manifest with this 
choice.  However, this choice has the advantage that $\xi^a$ and $\psi$ are related by
 $\xi^a=-\bar{\tilde\psi}\tilde\gamma^a \tilde\psi$ everywhere in the spacetime, 
not just in the asymptotic region.  We will see that with this choice the linearized integrands of the volume integral expressions for the charges $Q_\xi$ and $Q_\psi$ 
agree at each point.  A second natural choice for the extension of $\psi$ is to impose the Dirac-Witten equation (\ref{dirac}).  This choice makes the volume integrand
manifestly positive.  We will see, in this case, that the volume integrals for $Q_\xi$ and $Q_\psi$ again match, as they must.  However, the volume
integrands do not match on a point by point basis. 

The volume integrand for $Q_\xi$ is simply the normal component of the vector ${\cal C}^a$.   Let us start by specializing equation (\ref{volumecharge}) 
for ${\cal C}^a$ to the case of a deSitter CKV.  Given the properties of deSitter CKV's $\tilde\nabla_a\xi_b=f\tilde g_{ab}$, 
with $\tilde\nabla_a f=-{\Lambda\over 3}\xi_a$, it follows that
\be
\tnabla_b\tnabla_c\xi_d=-\tilde R_{cdb}^{\ \ \ e}\xi_e-  {\Lambda\over 3}(\tg_{cd}\xi_b +\tg_{bd}\xi_c -\tg_{bc}\xi_d).
\ee
Plugging this into equation (\ref{volumecharge}) for ${\cal C}^a$ then gives
\be\label{newcurrent}
{\cal C}^a = {\cal T}^{a b}\xi_ b- {\Lambda\over
3}K^{a\alpha b\beta}\tilde g_{\alpha\beta} \xi_ b
+f\left(\tilde\nabla_\beta K^{a\alpha b\beta}\right)\tilde g_{\alpha b},
\ee
which includes additional terms linear in $h_{ab}$,
 relative to the current ${\cal J}^a={\cal T}^{a b}\xi_b$ in the Killing vector construction.

We now follow the first approach and take $\psi$ to be a Killing spinor, satisfying 
\be
\hat{\tilde\nabla}_a\psi=0,
\ee
everywhere in the interior.  
From section (\ref{positiveenergy}), the volume integrand for $Q_\psi$ is the normal component of the conserved current ${\cal D}^b$ defined by the
right hand side of equation (\ref{spinoridentity})
\bea\label{newspinorcurrent}
{\cal D}^b= & \left (G^ b_{\  c}+\Lambda g^ b_{\  c}\right)
(-\bar\psi\gamma^ c\psi)
      +2 \overline{(\hat\nabla_a\psi)}
\gamma^{a b c}(\hat\nabla_ c\psi)  \\
&  +2iH\left ( \bar\psi\gamma^{a b}(\hat\nabla_a\psi)
      - \overline{(\hat\nabla_a\psi)}\gamma^{a b}\psi\right), \nonumber
\eea
The task is now to compare the currents ${\cal C}^a$ and ${\cal D}^a$ to linear order in $h_{ab}$, with the identification 
$\xi^a= - \bar{\tilde\psi} \tilde\gamma^a \tilde\psi$, where $\tilde\gamma^a$ are background gamma matrices satisfying 
$\{\tilde\gamma^a,\tilde\gamma^b\}=2\tilde g^{ab}$.  Making use of Einstein's equation, the first terms in equations (\ref{newcurrent}) 
and (\ref{newspinorcurrent}) clearly agree to this order.  To compare the other terms we need an expression for $\hat\nabla_a\psi$ to linear order,
which is found to be
\be
\hat\nabla_a\psi = -{1\over 4}(\tnabla_b h_{ac})\tilde\gamma^{bc} \psi +{i\over 4} Hh_{ab}\tilde\gamma^b\psi + {\cal O}(h^2).
\ee
The second term in equation (\ref{newspinorcurrent}) is second order in $\hat\nabla_a\psi$ and therefore does not contribute to linear order in
$h_{ab}$.  Dirac matrix algebra reduces the third term in (\ref{newspinorcurrent}) to 
\be\label{lastterm}
2iH\left ( \bar\psi\gamma^{a b}(\hat\nabla_a\psi)
      - \overline{(\hat\nabla_a\psi)}\gamma^{a b}\psi\right) =
      {\Lambda\over 3}(h^b_{\ c}-\tilde g^b_{\ c} h)(-\bar\psi\tilde\gamma^c\psi)
      -iH(\tnabla^b h-\tnabla_a h^{ab})\bar\psi\psi .
\ee
Making use of the relation for the conformal factor $f=iH\bar\psi\psi$ from equation (\ref{gradxi}) and the definition of $K^{abcd}$ in equation 
(\ref{superpotential}) then shows the equality of (\ref{lastterm}) with the last two terms in equation (\ref{newcurrent}) for ${\cal C}^a$.  
Therefore, we have shown that the volume integrands for the charges $Q_\xi$ and $Q_\psi$ agree pointwise to linear order,
\bea
Q_\psi = Q_\xi &=&\int_{\Sigma} \left ( T^{a b}n^a\xi_ b- {\Lambda\over
3}K^{ac bd}\tilde g_{cd} n^a\xi_ b
+f\left(\tilde\nabla_d K^{ac bd}\right)\tilde g_{c b}n^a\right ) \\
&=& \int_{\Sigma}\left ( T_{ab}(-\bar{\tilde\psi}\tilde\gamma^a \tilde\psi)n^b
-{iH\over 4\pi} \left(\tilde\psi^\dagger(\gamma^j\hat\nabla_j\tilde\psi)
- (\gamma^j\hat\nabla_j\tilde\psi)^\dagger\psi\right) \right ). \label{vectorvolume}
\eea

We now turn to the second alternative, extending $\psi$ to the interior by imposing the Dirac-Witten condition $\gamma^a\hat\nabla_a\psi=0$ everywhere.
The spinor field $\psi$ then has a first order correction,  
$\psi=\tilde\psi + \delta\psi$. The boundary term is unchanged.
 With the Dirac-Witten condition imposed, the volume expression
for $Q_\psi$ reduces to the manifestly positive expression in equation (\ref{positive}).  Further, because $\hat\nabla_i\psi$ is itself 
first order in $h_{ab}$, to linear order $Q_\psi$ given by the simple expression
\be\label{simple}
Q_\psi= \int_{\Sigma} T_{ab}(-\bar{\tilde\psi}\tilde\gamma^a \tilde\psi)n^b 
\ee
Consistency between equations (\ref{simple}) and (\ref{vectorvolume}) requires that
\be\label{linear}
\int_{\Sigma} \left(\tilde\psi^\dagger(\gamma^j\hat\nabla_j\tilde\psi)- 
(\gamma^j\hat\nabla_j\tilde\psi)^\dagger\psi\right) =0.
\ee
 To verify this, use
$\tilde\psi=\psi-\delta\psi$.  Then to linear order in $h_{ab}$  
$\tilde\psi^\dagger(\gamma^j\hat\nabla_j\tilde\psi)=\tnabla_j(\tilde\psi^\dagger\gamma^j\delta\psi)$.
The integrand in equation (\ref{linear}) is thus a total divergence 
and can be rewritten using Stokes theorem as
\be
{1\over 8\pi}\int_{\partial\Sigma} da_j(-2iH) \left( \tilde\psi^\dagger\gamma^j\delta\psi-\delta\psi^\dagger\gamma^j\tilde\psi\right).
\ee
Our earlier result from Section 3.2  $\gamma ^{\hat t}\delta \psi =-i\delta\psi$
 then implies that $\tilde\psi^\dagger\gamma^j\delta\psi=0$.
This shows that the terms in question do integrate to zero, and the different volume integrals
are equal.

This is interesting because the volume expresssion for the charge in (\ref{simple}) has a
simple physical meaning. Perturbatively, $Q_\psi$ is the integral of the
matter current $T_{ab}n^b$ in the direction of the CKV $\xi ^a$. Consider a comoving cosmological
observer, {\it i.e.} an observer whose four-velocity is the geodesic
$n^a=(\partial/ \partial t)^a$.  The perturbative mass density
measured by a comoving observer is
$\delta \rho  =T_{ab}n^a n^b$, and $\delta M =\int _V \delta\rho_{co}$ is the comoving
mass perturbation. To linear order in perturbation theory we then have
\be\label{linmass}
\delta Q_\zeta = a(t) \delta M_{co}.
\ee
The explicit factor of the scale factor comes from the CKV and the
choice of a comoving observer.

\subsection{Convergence of $Q_\psi$ and $Q_\xi$ for CKV's}

We have now established the equivalence of the two expressions for conserved charges associated with deSitter CKV's, $Q_\psi$ and $Q_\xi$. 
However, we have not yet studied the finiteness properties of these charges given the boundary conditions on asymptotically deSitter spacetimes
established in section (\ref{asymptotic}).  
It is simplest to work in terms of Hamiltonian variables. In section (\ref{asymptotic}),
we showed that appropriate falloff conditions on the metric and extrinsic curvature on a spatial slice $\Sigma$ are 
$q_{ij} -\tilde q_{ij} = {\cal O}(1/r)$ and $l^{ij}={\cal O}(1/r^3)$, where $K^{ij}=Hq^{ij}+l^{ij}$.
The $O(1/r)$ corrections to the extrinsic curvature enter only from the $Hq^{ij}$ term.  Since the boundary integral expression for
$Q_\xi$ is the same for KV's and CKV's we can treat both at once.  Let $\xi^a$ be a (C)KV and decompose $\xi^a$ as
$\xi^a = F n^a +\beta^a $, where $\beta^a n_a =0$.
The canonical momentum on the slice $\Sigma$ is given by
$\pi^{ij} =\sqrt {q} (K^{ij} -q^{ij} K )$.  Let $p ^{ij}=\pi^{ij} - \tilde\pi^{ij}$  be the perturbation to the canonical momentum and
$\rho_{ij}=q_{ij}-\tilde q_{ij}$ be the perturbation to the spatial metric.
The boundary term $Q_\xi$ as defined in equation (\ref{adm}) can then be rewritten 
in terms of these Hamiltonian variables as \cite{Abbott:1988ff}
\bea
 Q_\xi  &= & -{1\over 16\pi} \int da_i \left\{ \left( F(\tilde D^i \rho -\tilde D_j \rho 
^{ij}) -\rho \tilde D^i F +
\rho^{ij}\tilde D_j F \right)\right . \label{icv} \label{diverge}\\
&& \left .+  {1\over \sqrt {\rho} } \left( \beta ^i \tilde\pi ^{jk}\rho _{jk}
  -2 \beta ^j \tilde\pi ^{ik}\rho _{jk}-2\beta ^j p^i_{\ j} \right) \right\}\nonumber
\eea 
where $\tilde D$ is the covariant derivative operator on the spatial slice. 
For $Q_\xi$ to be finite, we see that $F$ must generically be independent of $r$, and that
$|\beta|$ fall off like $1/r$, as $r$ goes to infinity.  From equation (\ref{CKV}) we see that
 the only CKV for which 
$Q_\xi$ will be generically finite is the generator of conformal 
time translations $\zeta^a$ given in equation (\ref{conftime}). Further, we see that in general
none of the charges generated by the KV's are finite.  For example, the static time 
translation KV is given by $\chi^a = (\partial/\partial t)^a -H x^j (\partial/\partial x^j)^a $ 
in cosmological coordinates, which makes the momentum term in equation (\ref{diverge}) diverge. 
An exception to this is the Schwarzschild-deSitter spacetime itself.  In this case, 
substituting the spacetime (\ref{mcVittie}) into (\ref{diverge})  leads to a cancelation between the
momentum terms in equation (\ref{diverge}). Therefore the terms which depend on $\beta$ don't
contribute. The charge $Q_\chi$ then turns out to be finite and given simply by $Q_\chi=M$.

\subsection{Dynamical Interpretation of $Q_\xi$ for CKV's}

In the preceeding sections, we have explored properties of the conserved charge $Q_\zeta$ associated with
the conformal time translation CKV of deSitter and found that in certain respects it is a close parallel of the ADM mass for asymptotically
flat spacetimes.  In this section, we will briefly discuss another parallel, that $Q_\zeta$ is the value of the gravitational 
Hamiltonian if one evolves using a vector field that asymptotically approaches $\zeta$ at infinity.

In reference \cite{Regge:1974zd} Regge and Teitelboim studied the Hamiltonian evolution of asymptotically flat spacetimes. 
They showed that in order for the variational principle to be well defined, a boundary term must be added to the gravitational Hamiltonian.
On solutions to the equations of motion, the volume contribution to the Hamiltonian vanishes, and the value of the Hamiltonian is given by
the boundary term.  If the Hamiltonian evolution is carried out along a vector field that asymptotes to the time translation Killing vector
of flat spacetime, then this boundary term is simply the ADM mass.

It is straightforward to show that a similar situation holds for asymptotically deSitter spacetimes using the techniques of reference 
\cite{Traschen:1985bp}.  The gravitational Hamiltonian includes the volume term
$H_V= \int _V \sqrt{\gamma}(F{\cal H} +\beta ^a {\cal H}_a)$, 
where ${\cal H}$ is given by the left hand side of equation (\ref{hamiltonian}), 
$ {\cal H}_a = -2D_b\pi ^b _a/\sqrt{q} $, and $F$, $\beta ^a$ are Lagrange 
multipliers. Let the vector $\xi ^a = Fn^a+\beta ^a$ where $n^a$ is the
unit normal to $V$ and $n^a \beta_a =0$.   As discussed in reference \cite{Regge:1974zd}, the quantity $H_V$ is not precisely the 
correct functional to generate the Hamiltonian flow along the vector field $\xi$.  A 
boundary term $B_\xi$ must be added to the volume term to ensure that when one carries out the variation to derive
Hamilton's equations, the total boundary term which arises in the variation actually vanishes.  It was shown in
\cite{Traschen:1985bp} that finding the correct boundary term $B_\xi$ for a general
background spacetime amounts to computing the adjoint of the differential operator $(FH +\beta ^a H_a)$
and keeping track of the total derivative terms.  The expression found for $B_\xi$ in \cite{Traschen:1985bp} agrees with that given for 
$Q_\xi$ in equation (\ref{diverge}).  The equivalence of the expression in Hamiltonian variables and the covariant expression
in equation (\ref{adm}) was demonstrated in reference \cite{Abbott:1988ff}.
In the asymptotically deSitter case, choosing
$\xi^a$ to approach the conformal time translation killing vector $\zeta^a$ 
in the asymptotic region gives the Hamiltonian flow in conformal time.  On solutions to the equations of motion
$H_V$ vanishes, and the value of the  Hamiltonian is given by the
value of the boundary term $B_\zeta$, which is the same as our conserved charge $Q_\zeta$.

One might wonder about the feasibility of defining a mass with 
respect to  time translation in the static time coordinate,
which is a symmetry of deSitter. There turn out to be several interesting
drawbacks with
this approach. The static time KV is not timelike everywhere and one therefore looses
the positivity of the stress-energy term.  To get around this one can try setting the stress-energy to zero or
constraining it to be nonzero only within a horizon volume, where the KV is timelike. 
In Section (\ref{positiveenergy})  we showed that the vector fields $-\psibar \gamma ^{5}\gamma \psi$, 
where $\psi$ is a deSitter Killing spinor, are KV's rather than CKV's.  In Appendix A, we give the derivation of
a spinor identity in which the vector field which enters is this KV.
One finds that the spinor contribution to the volume term has 
positive definite and negative definite contributions, both of which are in general nonzero. 
Of course it still could be true that the sum is always positive, but the spinor construction does makes no indications of this.

There is another potentially interesting possibility which we have not
explored. Since the static KV is timelike  within a horizon volume,
one could study evaluating the boundary term  on the deSitter horizon. This would require that the spacetime
approach deSitter near the horizon in some suitable sense.
Although the spinor term in the KV construction is not sign definite, as discussed in
the previous paragraph, it is possible that this term is positive within a horizon volume.
Abbott and Deser \cite{Abbott:1982ff} showed that this is true to quadratic order.  Further analysis would be needed to
show that the gravitational contribution is positive to all orders. This choice of boundary is, of course,
observer dependent.  However, given the recent interest in the entropy of
deSitter spacetimes defined with respect to the part of the spacetime accessible to
an observer (see {\it e.g.} reference \cite{Bousso:2002fq}), investigations restricted to this portion of the spacetime might prove useful.

\appendix

\section{Spinor Identity for deSitter KV's \\ and anti-deSitter CKV's}

Generalize the Nester form of section (\ref{positiveenergy}) to
\be
B^{a b}=\bar\psi (\gamma^{\hat 5})^p\gamma^{a b c}\hat\nabla_ c\psi 
,
\ee
where $p$ can be $0$ or $1$.
One can then prove the more general identity
\bea
\nabla _a \left( B^{a b}  +  B^{\dagger a b}\right) n_ b  &= & \left( G_{a b}+\Lambda g_{a b}\right)
\xi^a n^ b \label{bigequation}\\  
&& + 2(-1)^p q^{ij}\hat\nabla_i\psi^\dagger (\gamma^{\hat
5})^p \hat\nabla_j\psi  -
2(-1)^p (\gamma^i\hat\nabla_i\psi)^\dagger(\gamma^{\hat
5})^p(\gamma^j\hat\nabla_j\psi) \nonumber\\ &&
+ (-1)^p H (-c+(-1)^p c^*)
\left[ \psi^\dagger(\gamma^{\hat 5})^p\gamma^j\hat\nabla_j\psi - \right . \nonumber \\
&& \left . (-1)^p(\gamma^j\hat\nabla_j\psi)^\dagger(\gamma^{\hat 5})^p\psi\right],\nonumber
\eea
where $\xi^a=-\bar\psi(\gamma^{\hat 5})^p\gamma^a\psi$ and $n^ b$ is
the unit normal vector to a spacelike slice.  When $\psi$ is a Killing spinor, then as discussed in section (\ref{basics})
for $p=0(1)$, $\xi^a$ a deSitter CKV (KV) or an anti-deSitter KV (CKV).
For $p=0$, equation (\ref{bigequation}) reduces to the result of section (\ref{positiveenergy}).
For $p=1$ the volume term, which is quadratic in $\hat\nabla\psi$, is easily seen to be
no longer sign definite.  Decompose $\psi$ into eigenvectors of $\gamma ^{\hat 5} $,
\be
\psi ={1\over 2}(1-\gamma ^{\hat 5} )\psi + {1\over 2}(1+\gamma 
^{\hat 5} )\psi
\equiv \psi _- +\psi _+ .
\ee
The second term on the right hand side is then given by
\be 
2 q^{ij}
(-\nabla_i \psi ^\dagger _+ \nabla _j \psi _+ + \nabla_i \psi 
^\dagger _- \nabla _j \psi _-) .
\ee
The Dirac-Witten equation written in terms of the projected spinors is given by
$\gamma ^j \nabla _j \psi _\mp + i {H\over 2} (n-1)\psi _\pm =0$, which has no solutions with $\psi _+ =0$ and nontrivial $\psi _-$.
From this we see that the volume integrand will include negative, as well as positive contributions.  It is still possible that the overall
result could be overall positive as argued in reference \cite{Torii:2001}, but the spinor construction does not appear to give any indication of this.

We now want to check that the charges $Q_\psi$ and $Q_\xi$ agree for the deSitter Killing vectors, as was the case for the charges corresponding
to the deSitter CKV's.  To do this, we start again by dividing the Nester form as in equation (\ref{dividebt}) and follow the analysis of 
the boundary term as in section (\ref{asymptotic}) above.  The terms which are linear in $H$, which canceled for $p=0$ , now add 
for $p=1$.  There is then an additional contribution to the boundary term in the KV case.
Using the perturbative result $\delta \gamma _a ={1\over 2} h_{ab}\tilde\gamma ^b$, the boundary
term becomes 
\be
Q_\psi  =   Q_\psi |_{p=0} +i {H\over 16\pi} 
\int ds_{ab}
\bar{\tilde\psi} \gamma^{\hat 5} ( h \tilde\gamma ^{ab} -\tilde\gamma ^{ac} h_c ^b 
+\tilde\gamma ^{bc} h_c ^a )\tilde\psi ,
\ee
where $ Q_\psi |_{p=0} $ is given in (\ref{finally}). The $p=0$ term
only depends on the spinor fields through the vector contribution
$\xi^a=-\bar{\tilde\psi}\tilde\gamma^a\tilde\psi$. The additional boundary term in the $p=1$ case 
depends on the product of two 
gamma matrices sandwiched between
$\bar{\tilde\psi}$ and $\tilde\psi$.  As one can see from equation (\ref{gradxitwo}), 
this quantity is proportional to the shear $w_{ab}=\nabla _{[a} \xi _{b]}$ of the Killing 
field $\xi^a$.  
% With $\hat\nabla _a \psi =0$, then for $p=1$, $w_{ab} \equiv \nabla _a \xi 
% _b =-i H \bar\psi \gamma^{\hat 5}
% \gamma _{ab} \psi$. 
The shear contribution of this shear term to $Q_\psi$ then matches the contribution to $Q_\xi$ coming from the 
shear term in equation (\ref{surfacecharge}).

\end{document}